\documentclass[slac_one]{revtex4}
\usepackage{graphicx}
\usepackage{fancyhdr}
\begin{document}

\title{Leptonic $B$ Decays at BaBar} %% Paper title goes here


\author{E. Baracchini (on behalf of the BaBar Collaboration)}
\affiliation{Universit\`a di Roma La Sapienza and INFN Rome, Roma, P.le A. Moro 2, Italy}

\begin{abstract}
We will present the most recent results on leptonic $B$ decays
 $B^{\pm(0)} \to K^{* \pm (0)} \nu \bar{\nu}$ and 
 $B^{\pm} \to \mu^{\pm} \nu$, based on the data collected by 
the BaBar detector~\cite{NIM} at PEP-II, an asymmetric
$e^+ e^-$ collider at the center of mass energy of the $\Upsilon(4S)$ resonance .
\end{abstract}

\maketitle

\thispagestyle{fancy}


\section{INTRODUCTION} % Section title should be in all capitals.
Rare $B$ decays have always been a standard probe for New Physics (NP)
searches. The very low Standard Model (SM) rate of these decays
often make them unaccessible with the present experimental datasets,
unless NP effects enhance the rate up to the current experimental
sensitivity.  Moreover, as NP effects can modify the decay kinematic,
particular attention must be payed in order to perform a model
independent analysis.

A $B$-Factory provides an unique environment where to investigate
these processes. The high number of $B \bar{B}$ pairs produced
by a $B$-Factory often allows to approach the needed experimental
sensitivity. Moreover, the clean environment and the closed kinematic of
the initial state enable to obtaining a 
very pure sample where to look for these decays.

In this work, we are going to present the most recent results 
in the searches of $B^{\pm(0)} \to K^{* \pm (0)} \nu \bar{\nu}$
and $B^{\pm} \to \mu^{\pm} \nu$, based on the data collected 
by the BaBar detector~\cite{NIM} at PEP-II, an asymmetric
$e^+ e^-$ collider operating at a center of mass energy of 
10.58 GeV, corresponding to the mass of the $\Upsilon(4S)$ resonance.

\section{ANALYSES OVERVIEW}

The common feature of the analyses presented in this work is the presence
of undetectable particles in the final state, the neutrinos $\nu$. This
particular characteristic calls for non-standard analysis techniques, 
which enable to deal with the lack of informations regarding these
particles. Typically, the closed kinematic of an $e^+ e^-$
collision is exploited to constraint through energy and four-vector
conservation the $B \bar{B}$ pairs, after both particles have been 
reconstructed.

Different approaches can be employed in the selection of the $B$
meson which is not decaying into the channel of interest ($B_{\rm{tag}}$): 
a totally inclusive reconstruction is applied on the $B_{\rm{tag}}$,
without trying to identify its decay product, whenever the additional
kinematic constraint coming from the two-body nature of the signal $B$
($B_{\rm{sig}}$) can be exploited, as in $B^{\pm} \to \mu^{\pm} \nu$.
The high efficiency obtainable with this method has as drawback
a poor energy resolution.
On the other hand, when more than one neutrino is present in the event, 
a recoil technique is needed: first, the $B_{\rm{tag}}$ is reconstructed
either in a semileptonic $B_{\rm{sl}} \to D^{(*)} l \nu$ or hadronic
$B_{\rm{had}} \to D Y$ ($Y = \pi,K $)
system. Then, the channel of interest is searched in the rest 
of the event (ROE), defined as the set of tracks and calorimeter
clusters not associated with the $B_{\rm{tag}}$. Both hadronic
and semileptonic recoil are employed for the 
$B^{\pm(0)} \to K^{* \pm (0)} \nu \bar{\nu}$ search. 
This method allows a very high resolution and purity, but has clearly
a low efficiency (1$\%$-0.1$\%$).

\section{$B^{\pm(0)} \to K^{* \pm (0)} \nu \bar{\nu}$ SEARCH}

\subsection{Theoretical Introduction}

In the SM $b \to s \nu \bar{\nu}$ processes occurs through FCNC and
are therefore forbidden at the tree-level. As these transitions
proceeds through one-loop box or electroweak penguin diagrams, they
are expected to be highly suppressed. In particular, due to the absence
of photon penguin contributions and long distance effects,
the $B^{\pm(0)} \to K^{* \pm (0)} \nu \bar{\nu}$ decay rate
can be calculated in the SM with less theoretical uncertainties
with respect  to the corresponding $b \to s l^+ l^-$.
The expected branching ratio (${\cal B}$) is
${\cal B}(B \to K \nu \bar{\nu}) = (1.3^{+0.4}_{-0.3}) \times 10^{-5}$~\cite{Buchalla:2000sk}.
However, this value can be enhanced in NP scenarios, where several
mechanism can contribute to the rate. For example, in Ref.~\cite{Buchalla:2000sk},
non-standard $Z^0$ couplings give rise to a contribution which
can bring and enhancement up to a factor 10. Moreover, new sources of
missing energy, such as light dark matter~\cite{Bird:2004ts}
or unparticles~\cite{Georgi:2007ek,Aliev:2007gr}, if accompanied
by a $K^{*}$, would contribute to the rate.

The kinematic of the decay can be described in terms of 
$s_{\nu \nu} = m^2_{\nu \nu}/m^2_B$, where $m_{\nu \nu}$ is the invariant
mass of the neutrinos pairs and $m_B$ is the $B$ meson mass.
As NP can strongly affect the decay in terms of the $s_{\nu \nu}$
shape~\cite{Buchalla:2000sk,Aliev:2007gr}, it is important
to not rely on any theoretical model when performing the analysis.

A previous search by the Belle Collaboration sets the upper limits (UL)
of ${\cal B} (B^{\pm} \to K^{* \pm } \nu \bar{\nu} < 1.4 \times 10^{-4}$
and ${\cal B} (B^{0} \to K^{* 0 } \nu \bar{\nu} < 3.4 \times 10^{-4}$~\cite{chen:2007zk}
The results presented here are the first completely
model-independent search for $B^{\pm(0)} \to K^{* \pm (0)} \nu \bar{\nu}$.

\subsection{$B^{\pm(0)} \to K^{* \pm (0)} \nu \bar{\nu}$ Analysis}

The $B^{\pm(0)} \to K^{* \pm (0)} \nu \bar{\nu}$ search is performed in the
recoil of both an hadronic (HAD) and a semileptonic (SL) system: the two different
tagging strategies provide non overlapping samples whose results can be 
combined as independent measurements. Moreover, the two
analyses have been developed in close synergy in order to combine
the final results more consistently as possible.  

The event selection start from the $B_{\rm{tag}}$ reconstruction:
for the SL analysis, neutral $D$ mesons are reconstructed in the
$K^- \pi^+$,$K^- \pi^+ \pi^0$,$K^- \pi^+ \pi^- \pi^+$ and $K^0_S \pi^+ \pi^-$
modes~\footnote{Charge conjugation is implied throughout this document, unless
explicitly stated.}. Charged $D$ mesons are reconstructed in the 
$K^- \pi^+ \pi^+$ and $K^0_S \pi^+$ final states. In the hadronic analysis,
the $B_{\rm{had}}$ is reconstructed in $B_{\rm{had}} \to D Y$ where
$Y = n\pi + m K + r K^0_S + q \pi^0$ with $n + m + r + q <6$ and $D$
is a generic charmed meson. About 1000 different decay chains are
considered. Charmed mesons are reconstructed in the same final states
used in the SL analysis, along with the additional channels
$D^+ \to K^+ \pi^- \pi^+ \pi^0, K^0_S \pi+ \pi^- \pi^+, K_S^0 \pi^+ \pi^0$.
For each reconstructed tagging $B$, a $K^{*}$ is searched in the ROE. A neutral
$K^{*}$ can be reconstructed in the $K^+ \pi^-$ mode, while a charged $K^{*}$ in
the $K^0_S \pi^+$ and $K^+ \pi^0$ modes. 

Considering that signal events have no
additional neutral particles produced in association with the $K^{*}$,
one of the most discriminating variable between signal and background is the
extra neutral energy $E_{\rm{extra}}$, defined as the sum of the energies of the 
electromagnetic calorimeter neutral clusters not used to reconstruct either
the $B_{\rm{tag}}$ of $B_{\rm{sig}}$ .

In the SL analysis, the signal yield
is extracted through a Maximum Likelihood (ML) fit to the final $E_{\rm{extra}}$
distribution, after selection criteria are applied to suppress continuum 
background. In HAD analysis, 
a loose selection is applied and all discriminating variables 
(including $E_{\rm{extra}}$) are used as inputs for a Neural Network (NN),
whose output variable $NN_{\rm{out}}$ is fitted to extract the number
of signal events.

The final selection efficiency for the SL analysis is $(5.6 \pm 0.7)\cdot 10^{-4}$ 
for $K^{*+} \to K^+ \pi^0$, $(4.3 \pm 0.6)\cdot 10^{-4}$ for $K^{*+} \to K^0_S \pi^+$ and 
$(6.9 \pm 0.8)\cdot 10^{-4}$ for $K^{*0} \to K^+ \pi^-$, while for the HAD analysis 
is $(6.7 \pm 0.6)\cdot 10^{-2}$ for $K^{*+} \to K^+ \pi^0$, 
$(6.1 \pm 0.7)\cdot 10^{-2}$ for $K^{*+} \to K^0_S \pi^+$ and
$(19.2 \pm 1.6)\cdot 10^{-2}$ for $K^{*0} \to K^+ \pi^-$.
The large difference is due to the fact that in the first case we normalize the 
${\cal B}$ measurement to the total number of produced $B \bar{B}$ pairs, while in the second
we use the number of reconstructed $B_{\rm{tag}}$.

The main systematics to the signal efficiency comes from the tagging
and the cut on the selection variables. The uncertainties
on the signal yield is mainly due to background distribution parameterization.
An uncertainty related to the residual model dependence is also taken into
account.

No significant signal is observed in the two analysis. A Bayesian approach
is employed to set upper limits (UL) at 90$\%$ of confidence level on the 
neutral (${\cal B}^0$) and charged (${\cal B}^{\pm}$) mode separately and on their combination.
The ULs are extracted from the two-dimensional posteriori PDF $P({\cal B}^{\pm},{\cal B}^{0})$,
where all the systematics are taken into account and the common ones are
assumed to be fully correlated. The combined UL extracted are

\begin{eqnarray}
{\cal B}(B^{\pm} \to K^{* \pm} \nu \bar{\nu}) &<& 8 \times 10^{-5} \nonumber\\
{\cal B}(B^{0} \to K^{* 0} \nu \bar{\nu}) &<& 12 \times 10^{-5} \nonumber\\
{\cal B}(B \to K^{* } \nu \bar{\nu}) &<& 8 \times 10^{-5} 
\end{eqnarray}

These results are more restrictive than previous measurements
from BaBar~\cite{Aubert:2004ws} and Belle~\cite{chen:2007zk}.

\section{$B^{\pm} \to \mu^{\pm} \nu $ SEARCH}

\subsection{Theoretical Introduction}

In the SM, the purely leptonic $B$ decays $B^{\pm} \to l^{\pm} \nu$
( $l = e, \mu, \tau$ ) 
proceed through the annihilation of the two quarks in the 
meson to form a virtual $W$ boson. The branching ratio can be cleanly 
calculated in the SM,    
\begin{equation}
{ \cal BR}(B^+ \rightarrow l^+ \nu_{l}) = \frac{G_{F}^{2} m_{B} m_{l}^{2}} {8\pi} 
 \biggl( 1- \frac{m_{l}^{2}}{m_{B}^{2}} \biggr)^{2} f_{B}^{2} |V_{ub}|^{2} 
 \tau_{B},
\end{equation}
where $G_F$ is the Fermi coupling constant, $m_{l}$ and $m_B$ are 
the lepton 
and $B$ meson masses, and $\tau_B$ is the $B^+$ lifetime. The decay rate 
is sensitive to the Cabibbo Kobayashi Maskawa 
matrix element $V_{ub}$ and the $B$ decay constant $f_{B}$ which describes 
the overlap of the quark wave functions within the meson. 
Currently, 
the uncertainty on $f_B$ is one of the main
factors limiting the determination of $V_{td}$ from precision $B^0\bar{B}^0$ 
mixing measurements. Given a measurement of 
$V_{ub}$ from semileptonic decays such as $B\rightarrow\pi l \nu$, 
$f_B$ could be extracted from  a measurement of the 
$B^{\pm} \to l^{\pm} \nu_{l}$ branching ratio. 

The SM estimate of the branching ratio for 
$B^{\pm} \to \tau^{\pm} \nu_{\tau}$ is $(1.59\pm0.40)\times 10^{-4}$ assuming 
$\tau_B$ = 1.638$\pm$0.011 ps, $V_{ub}$ = (4.39$\pm$0.33)$\times 10^{-3}$\cite{Barberio:2006bi} 
determined from inclusive charmless semileptonic $B$ decays and $f_B$ = 216 $\pm$ 22 MeV~\cite{Gray:2005ad}
from lattice QCD calculation. 
Due to helicity suppression, $B^{\pm} \rightarrow \mu^{\pm} \nu_{\mu}$ and $B^{\pm} \rightarrow e^{\pm} \nu_{e}$ are 
suppressed by factors of 225 and $10^{7}$ respectively, leading to   
branching ratios of ${\cal B}(B^{\pm} \rightarrow \mu^{\pm} \nu_{\mu} ) \simeq 4.7 \times 10^{-7}$ and
${\cal B}(B^{\pm} \rightarrow e^{\pm} \nu_{e}) \simeq 1.1 \times 10^{-11}$. 

Purely leptonic $B$ decays are sensitive to physics beyond the SM due to possible
insertion of New Physics (NP) heavy states in the annihilation process.
Charged Higgs boson effects may greatly enhance or suppress the 
branching ratio in certain two Higgs doublet models \cite{Hou:1992sy}. 
Moreover, as in annihilation processes the longitudinal component of the vector boson is directly 
involved, this decay allows a  direct test of Yukawa interactions in and beyond the SM. In particular, 
in a SUSY scenario at large $\tan \beta$ ($O(m_t/m_b) >> 1$), non-standard effects in helicity-suppressed 
charged current interactions are potentially observable, being strongly $\tan \beta$ dependent:
\begin{eqnarray}
{\cal B}(B^{\pm} \rightarrow l^{\pm} \nu_{l}) \approx {\cal B}(B^{\pm} \rightarrow l^{\pm} \nu_{l})_{\rm{SM}} \times \Big( 1-  \tan \beta^2 m_B^2/M_H^2 \Big)^2 .
\end{eqnarray}
These decays are also potential probes for Lepton Flavour Violation (LFV) in the ratios 
$R_{B}^{\mu/\tau}={\cal B}(B^{\pm} \rightarrow \mu^{\pm} \nu )/{\cal B}(B^{\pm} \rightarrow \tau^{\pm} \nu )$ and
$R_{B}^{e/\tau}={\cal B}(B^{\pm} \rightarrow e^{\pm} \nu )/{\cal B}(B^{\pm} \rightarrow \tau^{\pm} \nu )$~\cite{IP}.

\subsection{$B^{\pm} \to \mu^{\pm} \nu $ Analysis}

 $B^{\pm} \to \mu^{\pm} \nu_{\mu}$ is a two-body decay so the 
muon must be mono-energetic in the $B_{\rm{sig}}$ rest frame. The momentum $p^*$ of
the muon in the $B$ rest frame is given by
\begin{equation}
 p^* = \frac{m^2_B - m^2_{\mu}}{2m_B} \approx \frac{m_B}{2} \approx 2.46 {\rm  GeV}.
\end{equation}
where $m_B$ is the $B$ mass and $m_{\mu}$ is the muon mass.
At BaBar, the CM frame is a good approximation to the $B_{\rm{sig}}$ 
rest frame, so we initially select well-identified muon candidates with
momentum $p_{\rm{CM}}$ between 2.4 and 3.2 GeV/c in the CM frame. Since the neutrino produced in the 
signal decay is not detected, any other charged tracks or neutral deposits in a signal 
event must have been produced by the decay of the $B_{\rm{tag}}$. 
Once the $B_{\rm{tag}}$ is reconstructed from  the remaining visible energy in the event, 
we can refine the estimate of the muon momentum in the $B_{\rm{sig}}$ rest
frame ($p^*$). We use the momentum direction of the $B_{\rm{tag}}$ and assume a total momentum of 320 MeV/c in the CM frame (from the decay 
of the $\Upsilon(4S) \to B^+ B^-$) to boost the muon candidate into the reconstructed $B_{\rm{sig}}$ rest frame. 

Backgrounds may arise from any process producing charged tracks in the 
momentum range of the signal, particularly if the charged tracks are  
muons. The two most significant backgrounds are $B$ semileptonic decays involving
$b\rightarrow u\mu\nu_{\mu}$ transitions where the endpoint of the muon 
spectrum approaches that of the signal, and non-resonant $q \bar{q}$ (continuum) events 
where a charged pion 
is mistakenly identified as a muon.
Continuum backgrounds are suppressed using event shape variables, as the light-quark events
tend to produce a jet-like event topology as opposed to $B^+ B^-$ events which tend to be 
more isotropically distributed in space.  Several topological variables 
are combined in a Fisher discriminant~\cite{fisher}.

The two-body kinematics of this decay is now exploited by combining $p^*$ and $p_{\rm{CM}}$
in a second Fisher discriminant, whose output $p_{FIT}$ is used to extract the
number of signal events through a ML fit. The final selection efficiency is 4.64 $\pm$ 0.19 $\%$.

The main systematic to the measurement comes 
from the background PDF parameterization.

No signal is observed and a Bayesian approach is used to extract the UL

\begin{eqnarray}
 {\cal B}(B^{\pm} \to \mu^{\pm} \nu_{\mu}) < 1.3 \times 10^{-6}
\end{eqnarray}

at the 90\% confidence level. These results are more restrictive than previous measurements
from BaBar ~\cite{Aubert:2004yu} and Belle~\cite{Satoyama:2006xn}. 

\section{CONCLUSIONS}

The results presented in this work are on the final BaBar dataset, consisting
of about 426 fb$^{-1}$. The $B^{\pm(0)} \to K^{* \pm (0)} \nu \bar{\nu}$ search
is the first completely model independent result on these channel, which
does not rely on any theoretical assumption to perform the analysis. The 
upper limit on ${\cal B}(B^{\pm} \to K^{* \pm} \nu \bar{\nu})$,
${\cal B}(B^{0} \to K^{* 0} \nu \bar{\nu})$ (as well as the combined measurement)
and  ${\cal B} (B^{\pm} \to \mu^{\pm} \nu)$ are currently the most
stringent UL on these channels.

\begin{acknowledgments}
We would like to thank the organizers of the ICHEP 08 conference and
we are grateful for the excellent luminosity and machine conditions
provided by our PEP-II colleagues, 
and for the substantial dedicated effort from
the computing organizations that support BaBar.
\end{acknowledgments}

\end{document}